# Cooperative Proxy Servers Architecture for VoD to Achieve High QoS with Reduced Transmission Time and Cost

[1]M. Dakshayini[2] ,   Dr.T..R..Gopalakrishan Nair

[1]Research Scholar, Dr MGR University
Dept. of ISE, BMSCE, Bangalore
shantha_dakshu@yahoo.co.in
[2]Director, RIIC
D S Institutions,,Bangalore
trgnair@yahoo.com

***Abstract* - The aim of this paper is to propose a novel Voice On Demand (VoD) architecture and implementation of an efficient load sharing algorithm to achieve Quality of Service (QoS). This scheme reduces the transmission cost from the Centralized Multimedia Sever (CMS) to Proxy Servers (PS) by sharing the videos among the proxy servers of the Local Proxy Servers Group [LPSG] and among the neighboring LPSGs, which are interconnected in a ring fashion. This results in very low request rejection ratio, reduction in transmission time and cost, reduction of load on the CMS and high QoS for the users. Simulation results indicate acceptable initial startup latency, reduced transmission cost and time, load sharing among the proxy servers, among the LPSGs and between the CMS and the PS.**

## 1. INTRODUCTION

The tremendous growth of World Wide Web has resulted in an increase of bandwidth consumption throughout the internet. Proxy caching has been recognized as an effective technique to reduce network traffic. Caching is also an important mechanism for improving both the performance and operational cost of multimedia networks [10]. Recent web access patterns show frequent requests for a small number of popular objects at popular sites. So, a popular video can be transmitted to the same network link once per request. In the absence of caching, this approach results in server over load, network congestion, higher latency and even the possibility of rejection of the clients request. Caching proxy servers resolves the above issues by distributing the load across the network [3].

A VoD system usually has several servers and distributed clients over the entire network. These servers contain prerecorded videos and are streamed to the clients upon request from the clients. Proxy cache attempts to improve the performance of the overall network communication in three ways [9]. They are

i. Reduce the user perceived latency associated with obtaining documents as the proxy cache is normally close to the user.
ii. Lower the network traffic as the documents are available to the user and hence reduces the load on the network
iii. Reduce the service load

In the modern era, a number of caching and buffering techniques have been proposed to reduce the startup latency and bandwidth demand between the CMS and PS. Most of these techniques use proxy servers with large storage space for caching videos which are frequently requested. Cached data is serves the future requests and the uncached portions of the video are downloaded from the centralized servers [2].

Proxy servers have been widely used for multimedia contents to decrease the startup delay and to reduce the load of the centralized multimedia server. Recent works explores the advantages of connected proxy servers within the same LAN [3] [5] [8].

## 2. RELATED WORK

Authors, Tay and pang in [3] propose an algorithm called GWQ (Global waiting queue). This algorithm reduces the startup delay by sharing the load in a distributed loosely coupled VoD system by balancing the load between the lightly loaded proxy servers and heavily loaded proxy servers. They have replicated the videos evenly in all the servers. Hence, the storage capacity of individual proxy server should be very large to store all the videos. However, we propose to store the portion of the video only once across the proxies of the LPSG to accumulate more number of video blocks. Authors, Sonia Gonzalez, Navarro and Zapata in [4] propose a more realistic algorithm to share the load in a distributed VoD system. They suggest that their algorithm maintains a small startup delay using less storage capacity servers by allowing partial replication of the videos. They further store the locally requested videos in each server. Depending on the user requests, the popularity of the videos and percentage of replication is determined [2] and [7] considers the exchange of cached contents among the local proxy server without any coordinator. Our work differs in that we have made a group of proxy servers with a coordinator (Tracker) to share the videos among the proxy servers of LPSG more efficiently.

In this paper, we propose an efficient load sharing algorithm and a new VoD architecture for distributed VoD system. This architecture consists of a centralized multimedia server which is connected to a set of trackers. Each tracker is in turn connected to a group of proxy servers and these proxy servers are assumed to be interconnected using ring pattern. This arrangement is called as LPSG. Each of such LPSG, which is connected to CMS, is in turn connected to its left and right neighboring LPSGs in a ring fashion through its tracker.

Arranging the group of proxy servers in the form of LPSG provides several advantages:

- *Request-service delay*: Caching the most frequently asked videos across PSs of $L_p$ and sharing these videos among the PSs of $L_p$ can exploit the maximum bandwidth usage of the local area of networks.

- *Increased aggregate storage space*: By distributing the large number of videos across the PSs and TR of LPSG, high cache hit rate can be achieved. For example, if 10 PSs within a $L_p$ managed 500 Mbytes each, total space available is 5 Gbytes. 200 proxies of LPSG could store about 100 Gbytes of movies.

- *Load reduction*: Distribution of large number of videos on multiple proxies in LPSG provides service to more number of clients by passing downloading of the complete video from the main multimedia server.

- *Scalability*: By adding more number of PSs the capacity of the system can be expanded. Interconnected TRs increases the system throughput.

The further organization of the paper is as follows: Section 3 analyzes various parameters used in the problem. In section 4 we present a proposed approach and algorithm in detail. In section 5 we present a simulation model. Section 6 presents the simulation results and discussion. Section 7 summarizes the paper and refers to further work.

## 3. LOAD SHARING PROBLEM FORMULATION

This section describes the most important parameters and model considered in the evaluation process. Simulation details are given in the next section. By analyzing proxy server traces and content distribution network statistics, it becomes possible that the web objects are requested according to a Zipf-like distribution. The zipf law was originally formulated in terms of popularity of words in a text. It states that the frequency of i-th object is

$$p_i = \frac{k}{i} \quad i = 1,2,.....N \; where \; k = p_1 = \frac{1}{\Sigma_{i=1}^{N} i}$$

Zipf law is generalized as follows:

$$P_i = \frac{k}{i\alpha}$$

Where i assumes the value 0.986 for web object popularity.

**Table 1:** PARAMETERS IN THE MODEL

| Para | Definition |
|---|---|
| $S_i$ | Total length of $i^{th}$ video(minutes) |
| $N_v$ | Number of proxy servers in the system |
| N | Number of videos of the LPSG |
| $\lambda_i$ | Request rate for a specific movie in proxy server j(req/min) |
| TCOST | Transmission Cost for $(\{S_i- W_i\}^{CMS}$ |
| $Wt_i$ | Client's waiting time for a video $V_i$ |
| B | Total size of blocks at each PS |
| $W_i$ | Size of $(p-1)_i$ cached at $PS_q$, q=1..M |

With reference to fig. 1

Consider a group of N videos $V_1$, V2, ….. $V_N$ each of size(duration in minutes), $S_i$ with mean arrival rates $\lambda_1 \ldots \lambda_N$ respectively that are being streamed to the user using M proxy servers(PSs) $PS_1 \ldots PS_M$ of J LPSGs $L_p$, where p=1 . . .J. Each proxy server has a caching buffer large enough to cache total B minutes of video and k number of videos, W minutes of each video $V_i$ is $W_i$ i=1..N is cached across the PSs of $L_p$. That is M*B = $\sum_{i=1}^{N} Wi$ and B=$\sum_{i=1}^{K} Wi$. Let $b_i$ be the available bandwidth for $V_i$ between the proxy cache and CMS. After requesting for a video $V_i$, the streaming of that video will be delayed by $\{TT(S_i-W_i)^{CMS}b_i + TT(W_i)^{PS}\}$.let $\lambda_i > \lambda_j$ for $1 \le i < j \le N$ , i.e., popularity of these videos decreases gradually with the index so that $V_1$ and $V_N$ be the most and the least popular video respectively, i.e. minimum number of blocks will be cached for $V_1$ and large number of blocks will be cached for $V_N$. The total arrival rate for all videos is $\lambda = \sum_{i=1}^{N} \lambda i$. A user request a video $V_i$ with the probability $pi = \lambda i/\lambda$, i = 1 . . . N. Let the number of videos streamed from $L_p$ be N, leading to average user waiting time of $Wt_i(PS_q)$ and average transmission cost TCOST( ) where $Wt_i$( ) and TCOST() are non-linear functions, i= 1 … N. The optimization problem to maximize the streaming of number of blocks of videos from $L_p$ by sharing the videos among the PSs of LPSG $L_p$ so as to minimize the average transmission cost TCOST and user waiting time Wt for a video i at a PS q can be formulated as follows:

$$\text{Minimize } \sum_{i=1}^{N} \{Si - Wi\}$$
$$\text{Minimize TCOST } (\{S_i- W_i\}^{CMS} ),$$
$$\text{Minimize } Wt_i(PS_q)$$

$$\text{Subject to } M*B = \sum_{i=1}^{N} Wi \;,\quad B=\sum_{i=1}^{K} Wi \quad \text{and } Wi>0$$

## 4. PROPOSED ARCHITECTURE AND ALGORITHM

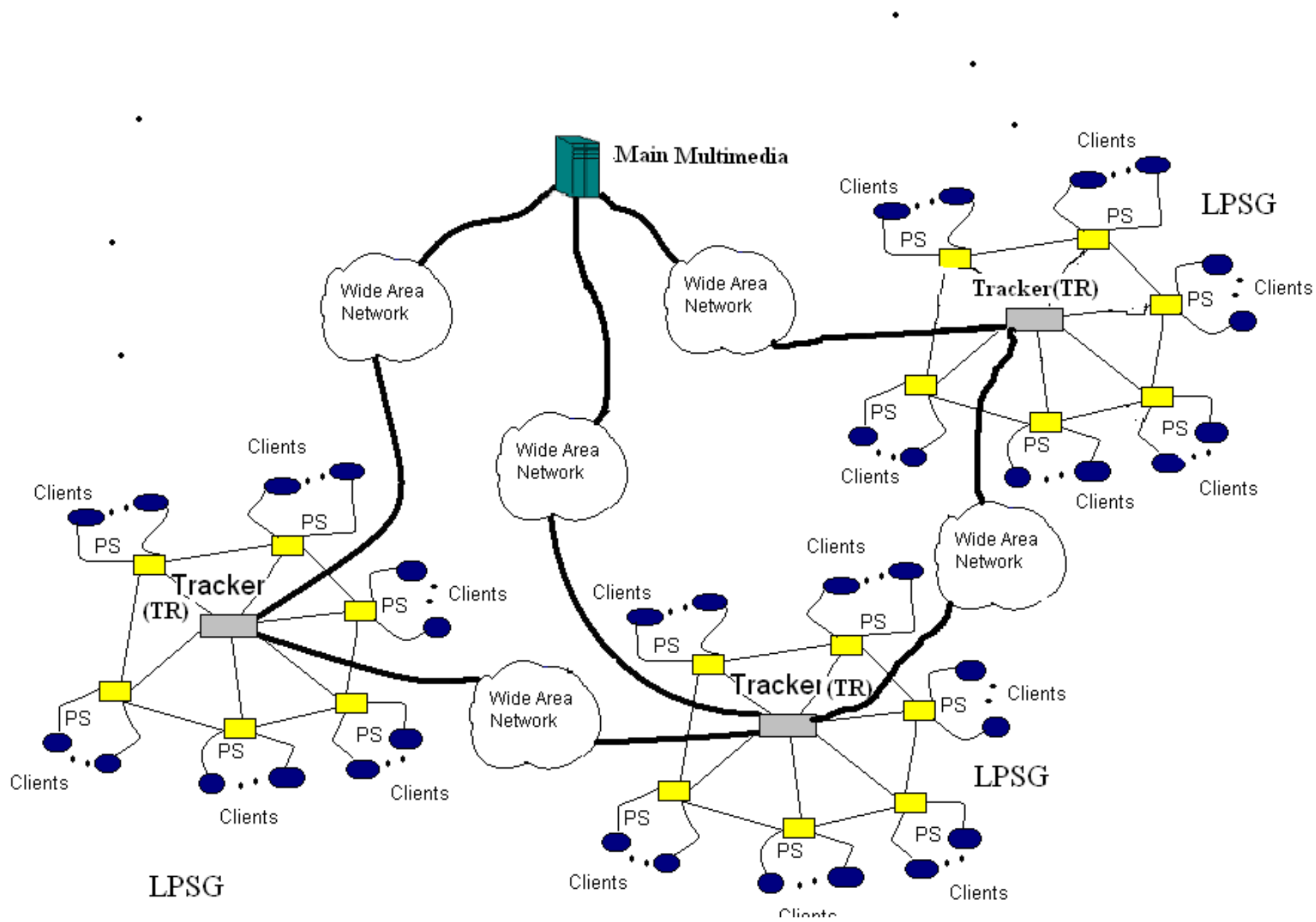

**Figure 1.** Cooperative Proxy servers Architecture

### 4.1 Architecture

The proposed VoD architecture is as shown Fig.1. Our proposed architecture consists of a CMS, which is connected to a group of trackers (TRs) where each TR is in turn connected to a set of PSs. These PSs are connected among themselves in a ring fashion. And also to each of these PSs a large number of users are connected [LPSG]. All these LPSGs are interconnected through their TR in a ring pattern. The TR caches some portion of video $V_i$ where i=1…N according to the frequency of requests for video Vi at any one PS of $L_p$ only once. Then PS transmits cached data to the client through LAN using its less expensive bandwidth. The TR is also a PS with high computation power, which coordinates and maintains a database that contains the information of the videos present in each PS in that $L_p$ and also free cache space at each PS.

The CMS, the TR and the PSs of LPSG are assumed to be interconnected through high capacity optic fiber cables. In the beginning, all the videos are stored in the CMS. TR, caches the portion $W_i$ of the video $V_i$ at any one of the proxy PSq of Lp only once, as it streams the video $V_i$ from CMS to the client upon the request. It decides the Size (number of blocks (minutes)) of this portion $W_i$ of $V_i$ to be cached in proportion to its popularity i.e size(Wi) α pop($V_i$). So, more number of blocks will be stored for most popular video. This setting drastically reduces the frequent access of CMS by serving maximum number of clients with maximum Wi from $L_p$. We assume that the frequency of requests to a video follows Zipf law of distribution and bandwidth $b_i$ will be available for $(S_i - W_i)^{CMS}$.

### 4.2 Optimal Load sharing Algorithm

In this section, we explain how our load sharing algorithm handles the user request of various scenarios as follows.

Case 1: If the $V_{req}$ is present in $PS_q$, then the video will be streamed to the user immediately and hence the initial startup latency is very low, and TCOST ($\{S_i - W_i\}^{CMS)}$ is very low.

Case 2: If the $V_{req}$ is not present in the $PS_q$, then the same is intimated to the TR. The TR checks whether this $V_{req}$ is present in any of the PS in that $L_p$. If the $V_{req}$ is present in one of the PSs in that $L_p$, then the TR checks whether the PS in which the $V_{req}$ was found is neighbor to the requested $PS_q$ [NBR ($PS_q$)]. If so, the TR($L_p$) initiates the streaming of the $V_{req}$ from that NBR($PS_q$) to the requested $PS_q$ and the same is intimated to the requested $PS_q$ and hence the Wt($V_{req}$) and TCOST($\{S_i - W_i\}^{CMS}$ ) is very small.

Case 3: The $V_{req}$ may not be present in the $PS_q$ and also it may not be present in the NBR($PS_q$) but may be present in the PS other than the NBR($PS_q$). Then the TR ($L_p$) intimates the same to $PS_q$ and initiates the streaming of the $V_{req}$ from this PS to the requested $PS_q$ through the optimal path found by the TR and hence the Wt ($V_{req}$) and TCOST ($\{S_i - W_i\}^{CMS)}$ is relatively higher, but acceptable with high QoS.

Case 4: If the $V_{req}$ is not present in any of the PSs in that LPSG, then the TR ($L_p$) Passes the request to the tracker of NBR ($L_p$). That is TR (NBR ($L_p$)) checks whether the $V_{req}$ is present in any of the PS in its LPSG using perfect hashing. If the $V_{req}$ is present in any of the PSs in that LPSG, then the TR(NBR($L_p$)) selects the optimal streaming path from the selected PS(NBR($L_p$)) to the requested $PS_q$ and intimates the same to TR($L_p$) and then initiates the streaming of the $V_{req}$ from the selected PS(NBR($L_p$)) to the TR($L_p$) through this optimal path. Then the TR ($L_p$) in turn continue the streaming of $V_{req}$ to the requested $PS_q$ and the same is intimated to the requested $PS_q$ and hence the Wt ($V_{req}$) and TCOST ($\{S_i - W_i\}^{CMS) is}$ relatively higher, but acceptable with high QoS.

Case 5: If the $V_{req}$ is not present in any of the PSs of its NBR($L_p$) also, then the TR($L_p$) decides to download the $V_{req}$ from CMS and stream to $PS_q$. So Wt($V_{req}$) and TCOST($\{S_i - W_i\}^{CMS}$ ) is maximum but very few requests are served from CMS.

Whenever the sufficient buffer and bandwidth is not available in the above operation the user request is rejected, which is a very rare possibility as shown by our simulation results.

**2.3 Simulation Algorithm:**

When a request for a video V ($V_{req}$) arrives at a particular time t to $PS_q$ of $L_p$ do the following:

```
If (V_req is present at PS_q)
  {
    Stream the V_req to the user immediately from  PS_q
  }
else
  {
    Pass  request to TR(L_p)-perform perfect Hashing
      If (V_req Є PS(L_p) )
        {
          If (PS(L_p) is left or right neighbor to  PS_q)
            {
              TR(L_p) initiates the Streaming of the
              V_req  to the user from  NBR(PS_q)
              through PS_q
            }
          else
            {
              TR(L_p)  streams the V_req to the user
              through PS_q from PS(L_p) using
              the optimal path found.
            }
        }
      else
        {
          Pass the request to left or right NBR (L_p)

        If (V_req Є PS(NBR(L_p)))
            {
              TR(NBR(L_p)) Streams of the
              V_req  to the user from  PS(NBR(L_p))
              through TR(L_p) and PS_q using the optimal path found
            }
        else
            {
              TR(L_p)  downloads and streams the
              V_req to the user through PS_q from CMS
```

```
        }
    }
}
```

## 5. SIMULATION MODEL

Our simulation model consists of a single CMS and a set of 6 TRs. All these TRs are interconnected among themselves in a ring fashion. Each of these TR is in turn connected to a set of 6 PSs. These PSs are also interconnected among themselves in a ring fashion. We use the video hit ratio (VHR) to measure the performance of our proposed approach more correctly. In addition we also use the average client waiting time, number of access to main server and the network transmission cost as performance metrics.

We assume that the request distribution of the videos follows a zipf-like distribution and transmission delay between the proxy and the client as 100ms, transmission delay between proxy to proxy as 200ms, transmission delay between TR and PS as also 200ms, transmission delay between the main server and the proxy as 1200ms, transmission delay between tracker to tracker 300ms, and the size of the cached video as 280MB to 1120MB (25min – 1hr) in proportion to its popularity.

## 6. SIMULATION RESULTS

The simulation results presented below are an average of several simulations conducted on the model. Consider Fig.2, which shows the average number of blocks of videos served from various proxies of LPSG and NBR ($L_p$), and the average number of blocks of videos served from CMS which is only about 39%.

As the maximum amount of most frequently asked videos have been cached and streamed from the LPSG with cooperation of PSs, and the coordination of TR of LPSG, Our scheme increases the video hit ratio and reduces the number of access to CMS by 30% -40% as shown in Fig 4 and Fig.5. So the network traffic along the server--client path reduces, in turn transmission cost and transmission time is also reduced significantly when compared to the system with single proxy as shown in Fig 3. More number of blocks of frequently requested videos are cached and shared among the proxies of Lp. So when there is a request for any of these $i^{th}$ video, streaming starts from one of the PS immediately and hence the Wt (Vi) and TCOST ({Si- Wi} CMS) is very less.

If the requested videos are present at NBR (PSq) of Lp, then these videos are streamed from NBR (PSq) to the client through PSq, so the clients Wt for these videos is very small. And if the requested videos are present in OTR (PSq), then these Videos are streamed from OTR (PSq) to the client through PSq, so the client’s Wt for these videos is relatively higher, Otherwise also, some good number of videos are served from NBR (Lp), which reduces frequent downloading of requested videos from CMS to the PSq which in turn reduces the initial playout delay at the clients for the requested videos which are not present at PSq.

And a very few numbers of blocks of videos are streamed from the CMS, when the Vreq is neither present in that Lp, nor in NBR (Lp). Even though the initial startup delay and transmission cost seems to be more as shown in fig.6, it is acceptable because on an average 60%-80% of the videos are cached and streamed from Lp and NBR (Lp) by assuring high QoS as shown in Fig.2 and Fig.3 and only 40 %-20% of the videos are downloaded from CMS which reduced the number of access to CMS and hence the transmission cost and time reduced significantly.

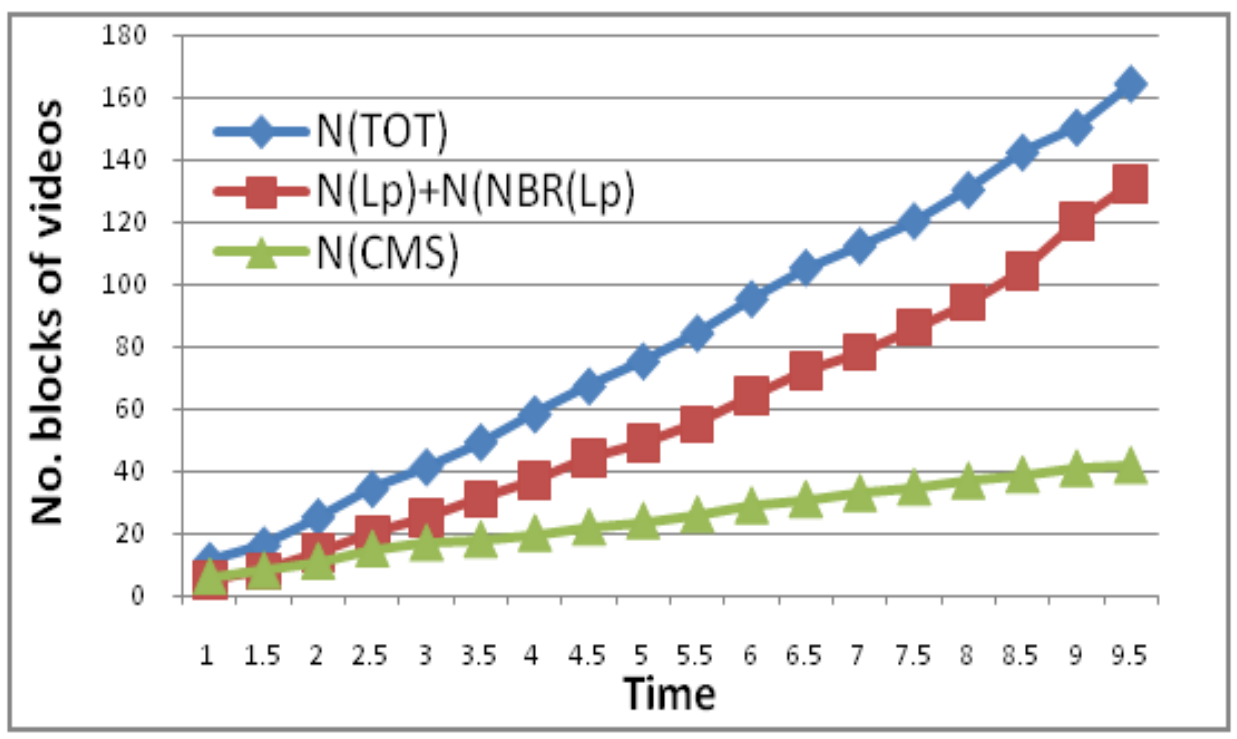

**Figure. 2** No. of Blocks of videos streamed from Lp+NBR(Lp), CMS Vs Time

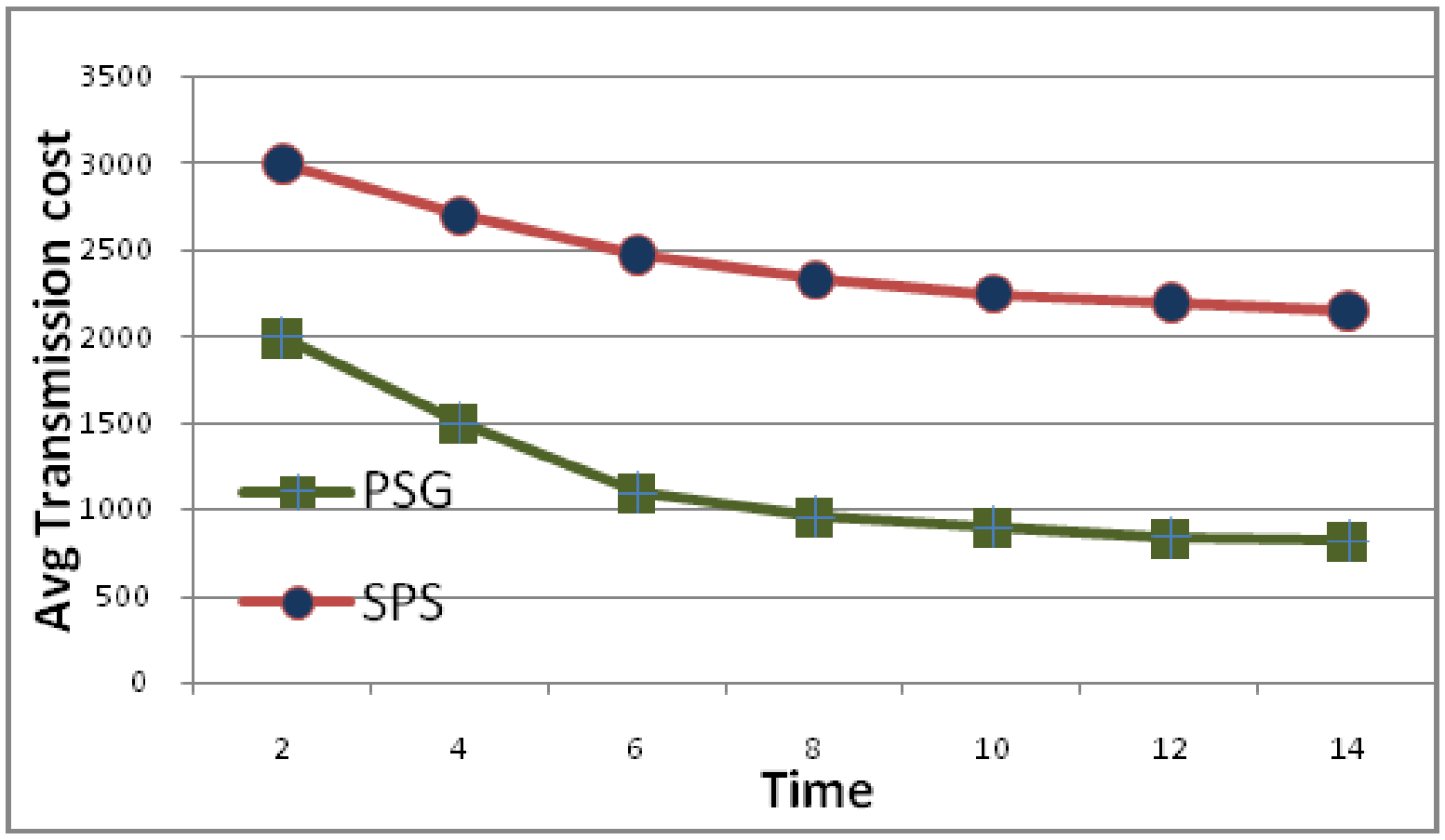

**Figure. 3** Average transmission cost with PSG Vs SPS

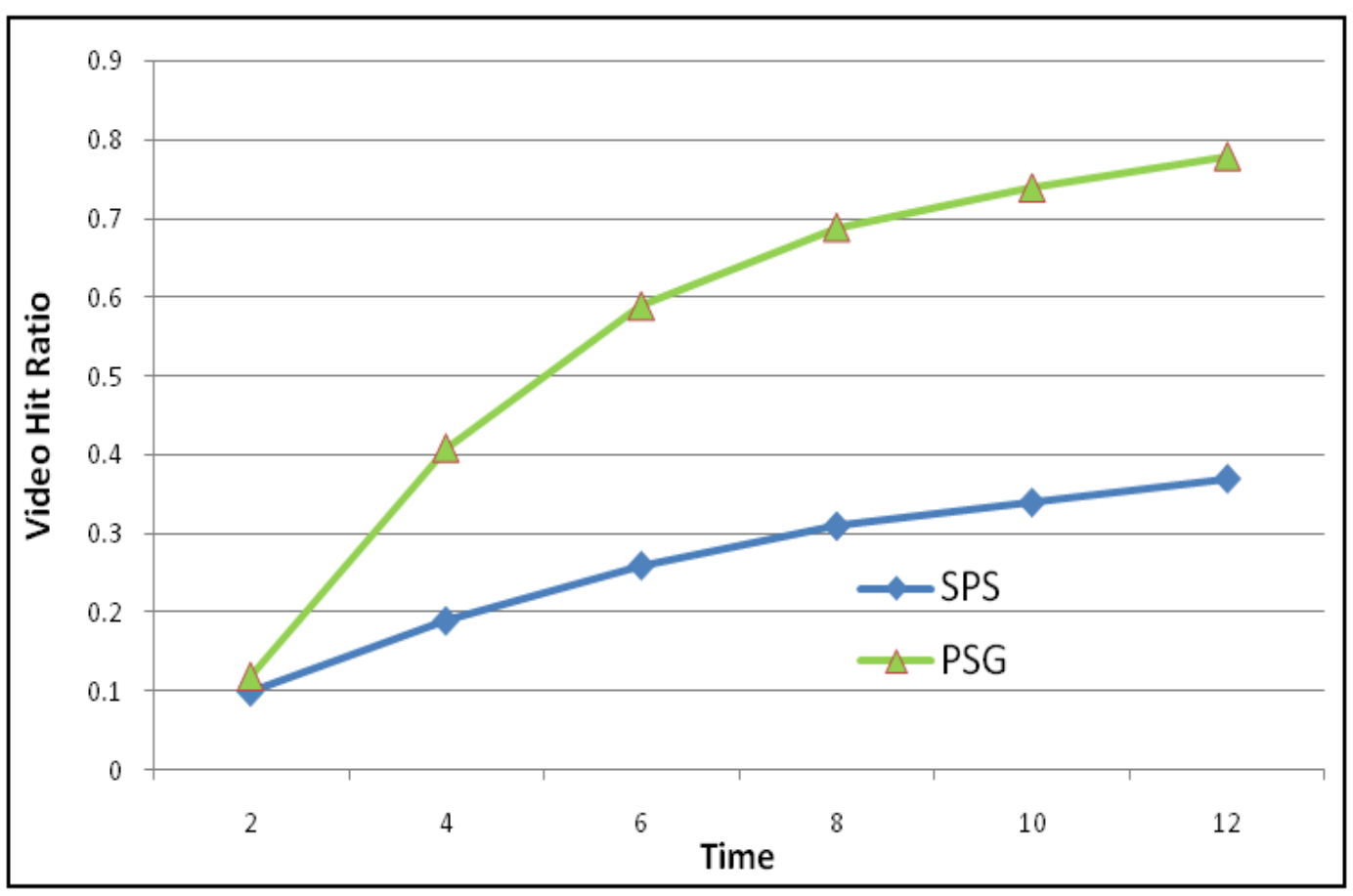

**Figure. 4** Video Hit Ratio with PSG Vs SPS

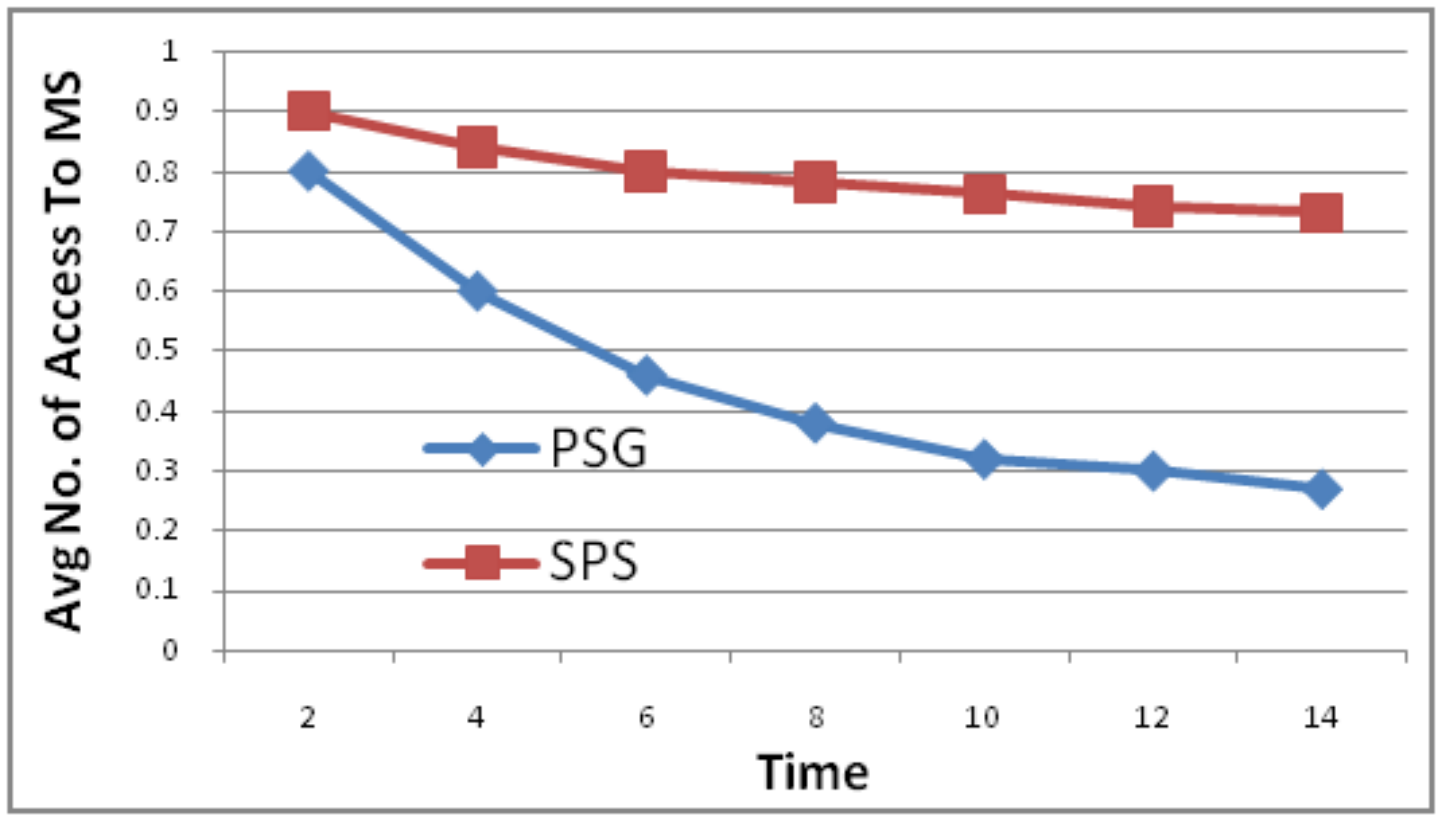

**Figure 5.** Average No. of Access to MS with PSG Vs SPS

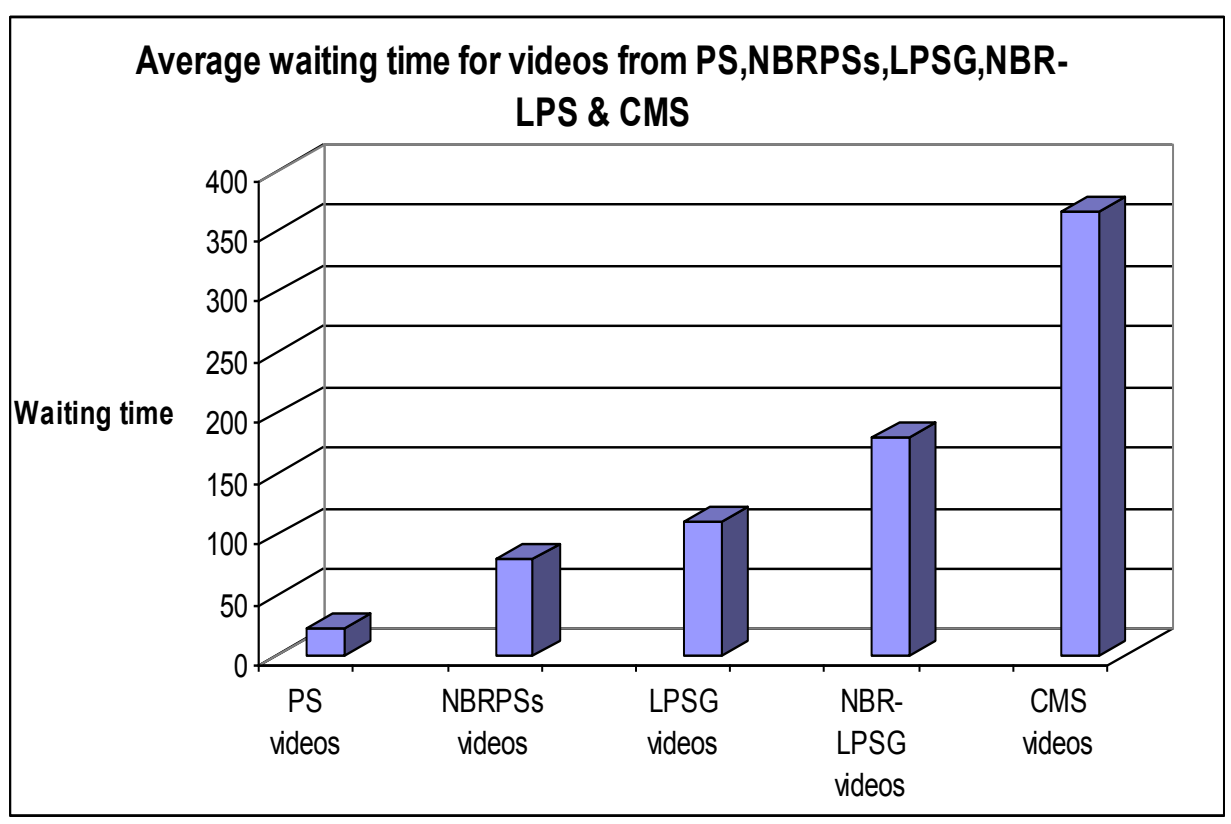

**Figure 6.** Average waiting time for video from PS,NBR(PS), Lp,NBR(Lp) and MS

## 7. CONCLUSION

In this paper we have proposed an efficient video sharing mechanism and an architecture where proxies cooperate with each other to achieve reduced transmission cost and initial start up delay, by caching and streaming maximum portion of the most frequently requested videos among the proxies of $L_p$. Our simulation results demonstrated that our proposed approach has reduced the average network transmission cost of the system, initial startup delay for the videos requested at $PS_q$, and also the load of CMS by caching maximum portion of the most popular videos in proportion to their popularity across the PSs of $L_p$. And sharing of these videos among the proxies of the system also reduces the server-to-client transmission cost and time, maintains high QoS for the users and has removed the video redundancy among the proxy servers. The future work is being carried out to improve the performance using dynamic buffer management at $PS_q$.